\documentclass[12pt]{article}

\usepackage{jheppub}
\usepackage{amsmath,amssymb}
\usepackage[font=small,font=sf,margin=1cm]{caption}
\usepackage{subcaption}
\usepackage[per-mode=symbol]{siunitx}
\usepackage{bm}
\usepackage{soul}
\usepackage{todonotes}
\usepackage[toc,page]{appendix}
\usepackage{tocloft}
\usepackage[autostyle=true]{csquotes}

\usepackage{cleveref}

\usepackage{parskip}
\newtheorem{definition}{Definition}[section]
\newtheorem{theorem}{Theorem}[section]

\newtheorem{example}{Example}

\newcommand\be{\begin{equation}}
\newcommand\ee{\end{equation}}
\newcommand\bea{\begin{eqnarray}}
\newcommand\eea{\end{eqnarray}}

\renewcommand\tilde{\widetilde}

\title{\begin{center}Generating Triangulations and Fibrations \\ with Reinforcement Learning\end{center}}
	
\author[a]{Per Berglund,}
\author[a]{Giorgi Butbaia,}

\author[b,c,d,e]{Yang-Hui He,}
\author[b,c]{Elli Heyes,}
\author[f]{Edward Hirst,}
\author[g]{Vishnu Jejjala}

\affiliation[a]{Department of Physics and Astronomy, University of New Hampshire, Durham, NH 03824, USA}
\affiliation[b]{London Institute for Mathematical Sciences, Royal Institution, London, W1S 4BS, UK}
\affiliation[c]{Department of Mathematics, City, University of London, EC1V 0HB, UK}
\affiliation[d]{Merton College, University of Oxford, OX1 4JD, UK}
\affiliation[e]{School of Physics, NanKai University, Tianjin, 300071, P.R.\ China}
\affiliation[f]{Centre for Theoretical Physics, 
Queen Mary University of London, E1 4NS, UK}
\affiliation[g]{Mandelstam Institute for Theoretical Physics, School of Physics, NITheCS, and CoE-MaSS, University of the Witwatersrand, Johannesburg, WITS 2050, South Africa\\}

\emailAdd{per.berglund@unh.edu}
\emailAdd{giorgi.butbaia@unh.edu}
\emailAdd{hey@maths.ox.ac.uk}
\emailAdd{elli.heyes@city.ac.uk}
\emailAdd{e.hirst@qmul.ac.uk}
\emailAdd{v.jejjala@wits.ac.za}

\preprint{\begin{flushright}
QMUL-PH-24-10
\end{flushright}}

\abstract{
We apply reinforcement learning (RL) to generate fine regular star triangulations of reflexive polytopes, that give rise to smooth Calabi-Yau (CY) hypersurfaces. We demonstrate that, by simple modifications to the data encoding and reward function, one can search for CYs that satisfy a set of desirable string compactification conditions. For instance, we show that our RL algorithm can generate triangulations together with holomorphic vector bundles that satisfy anomaly cancellation and poly-stability conditions in heterotic compactification. Furthermore, we show that our algorithm can be used to search for reflexive subpolytopes together with compatible triangulations that define fibration structures of the CYs.
}

\begin{document}
\maketitle

\section{Introduction}
There are many possible top down paths to low energy theories like the Standard Model.
Typical ingredients for obtaining four-dimensional effective field theories starting from ten-dimensional superstring theory include: i) a Calabi--Yau (CY) threefold, ii) orientifold planes, iii) fluxes, and iv) D-branes with open strings.
The landscape of possible effective field theories is vast, in part due to the large number of CY geometries, and we do not yet fully understand the principles that underlie a solution to the vacuum selection problem.
One obstacle is that the number of explicit models is still rather small, and isolated to particular corners of the space of theories.
In this work, we investigate the first building block (i) and use machine learning to propose new CY manifolds for string compactification.

The largest known class of compact CY threefolds is constructed as hypersurfaces in toric varieties:
any fine regular star triangulation (FRST) of a four-dimensional reflexive polytope defines a toric Fano variety in which any generic anticanonical hypersurface is a smooth CY threefold~\cite{batyrev1993dual,batyrev1994calabiyau}.
Each such triangulation gives rise to a different ambient toric variety and potentially different CY hypersurfaces.
However, triangulations of two different polytopes can also produce equivalent CY hypersurfaces.

In 2000, Kreuzer and Skarke (KS) constructed all $473,800,776$ four-dimensional reflexive polytopes.
The CY manifolds inherit their topological properties from the polytope, giving rise to threefolds with $30,108$ distinct Hodge diamonds~\cite{kreuzer2000complete}.\footnote{The statistics of polytopes is discussed in~\cite{He:2015fif}, with supervised machine learning in~\cite{Bao:2021ofk}.}
As we lack a complete classification of all FRSTs of the polytopes, we do not know how many  different CY geometries there are. 
The total number of FRSTs of polytopes in the KS list has been estimated to be as large as $1.53\times 10^{928}$~\cite{demirtas2020bounding}, though the number of topologically inequivalent CY threefolds is known to be much smaller than this.
Generalizing the discussion in~\cite{Hubsch:1992nu}, there has been revived interest in calculating diffeomorphism classes of CY threefolds~\cite{Jejjala:2022lxh, Chandra:2023afu, Gendler:2023ujl}.
A consequence of Wall's theorem~\cite{wall} is that the CY hypersurfaces coming from two different FRSTs of the same reflexive polytope $\Delta$ that have the same restriction on the two-faces of $\Delta$ are topologically equivalent.
It has been estimated that the number of non two-face equivalent FRSTs of four-dimensional reflexive polytopes is $<1.65\times 10^{428}$~\cite{demirtas2020bounding}.
Both the upper bounds  
are dominated by the single largest polytope with $496$ lattice points and which produces CY threefolds with $h^{1,1}=491$.\footnote
{
The numbers obtained from a na\"{\i}ve counting of effective field theories become still more astronomic in F-theory when we consider toric CY fourfolds~\cite{Taylor:2015xtz}.}
For low values of $h^{1,1}$, it is possible to exhaustively compute FRSTs of four-dimensional reflexive polytopes~\cite{Altman:2014bfa}, but it is clearly infeasible to sustain this brute force attack once the number of K\"ahler moduli increases.
We therefore seek a method of efficiently and effectively sampling triangulations. 

With multiple criteria, finding FRSTs is an example of a multiobjective optimisation problem (MOOP).
Regression is unsuitable for this type of problem since any loss function will have local minima, where triangulations satisfy some but not all of the desired criteria.
Popular methods for solving MOOPs are genetic algorithms (GAs) and reinforcement learning (RL), which explore an ``environment" in order to maximize a fitness or reward function. 
GAs have  been shown to generate reflexive polytopes~~\cite{Berglund:2023ztk}, and RL has achieved successes for string model building in other contexts~~\cite{Halverson:2019tkf,Harvey:2021oue,Constantin:2021for,Abel:2021rrj}.
Here, we extend this work to obtain FRSTs of reflexive polytopes with RL.\footnote{
We initially tried using GAs to approach this problem, but found RL to be better suited to the task.
Parallel work in~\cite{MacFadden:2024him} constructs FRSTs using GA methods.}

The organization of this paper is as follows.
In Section~\ref{sec:model}, we present the RL model used in the analysis.
In Section~\ref{sec:results}, we demonstrate that we can obtain FRSTs of reflexive polytopes for low $h^{1,1}$. 
We perform targeted searches for CY geometries whose intersection numbers are conducive to string model building and use our algorithm to find fibration structures of the CYs.
In Section~\ref{sec:outlook}, we outline the next steps and conclude.
Appendix~\ref{sec:toric} recalls relevant features of toric geometry.
Appendix~\ref{sec:RL} reviews RL.

Our code, is available on \href{https://github.com/elliheyes/Triangulation-Generation}{\textsf{GitHub}}~\cite{github}.

\section{The model}\label{sec:model}

In our investigations, we apply deep Q-learning, a model-free reinforcement learning (RL) algorithm involving
an optimal Q-function $Q:\mathcal{S}\times\mathcal{A}\rightarrow\mathbb{R}$ which gives the value of action $a\in\mathcal{A}$ in a particular state $s\in\mathcal{S}$. 
In short, given a state $s_{t}$, the Q-learning algorithm selects an action $a_{t}$, computes the reward and moves to the next state $s_{t+1}$.
The Q-function is then updated based on the Bellman equation~\eqref{eq:bellman}.
In deep Q-learning as opposed to regular Q-learning the Q-function is represented by a deep neural network.  
Additional details about RL and Q-learning are in Appendix~\ref{sec:RL}. 

The neural network used for the Q-function consists of two hidden layers of $100$ and $200$ neurons using ReLU activation function.
The loss function used was mean squared error and the network was trained using Adam optimisation. 
In all our investigations the learning rate $\alpha$ and discount factor $\gamma$ in~\eqref{eq:bellman} were set to $0.7$ and $0.618$, respectively.

Wall's theorem~\cite{wall} establishes that two CY threefolds are topologically equivalent if they share the following invariants: 
i) Hodge numbers $h^{1,1}$ and $h^{1,2}$,
ii) first Pontrjagin class $\mathrm{p}_1$, or equivalently, the second Chern class $c_{2}$,
and iii) triple intersection numbers $d_{ijk}$.
The second Chern class and the triple intersection numbers are defined up to $SL(h^{1,1}, \mathbb{Z})$ equivalence.

The Hodge numbers of CY threefolds constructed as hypersurfaces in toric varieties built from four-dimensional reflexive polytopes can be obtained from the polytope data alone.
The second Chern class and the triple intersection numbers, however, depend on the triangulation of the two-faces.
Thus, two FRSTs $\mathcal{T}_{1}$ and $\mathcal{T}_{2}$ of the same reflexive polytope $\Delta$ which give the same two-face restriction give rise to topologically equivalent CY threefolds~\cite{Altman:2014bfa, wall}.
With this in mind, we encode triangulations of the polytope $\Delta$ by their two-face restrictions.
We first compute all the two-faces of $\Delta$: $\{F_{1},...,F_{N}\}$ and choose an ordering for these faces.
Then for each two-face $F_{i}$, we compute all fine, regular triangulations $\{\mathcal{T}^{i}_{1},...,T^{i}_{M_{i}}\}$ and decide on an ordering for these triangulations.
We then define a state $s\in\mathcal{S}$ as a collection of two-face triangulations given by the two-dimensional bitlist $s^{i}_{j}$ where $s^{i}_{j}=1$ if the $j$-th triangulation $\mathcal{T}^{i}_{j}$ of $F^{i}$ is chosen and $0$ otherwise. 

\begin{figure}[h]
    \centering
    \includegraphics[width=0.6\textwidth]{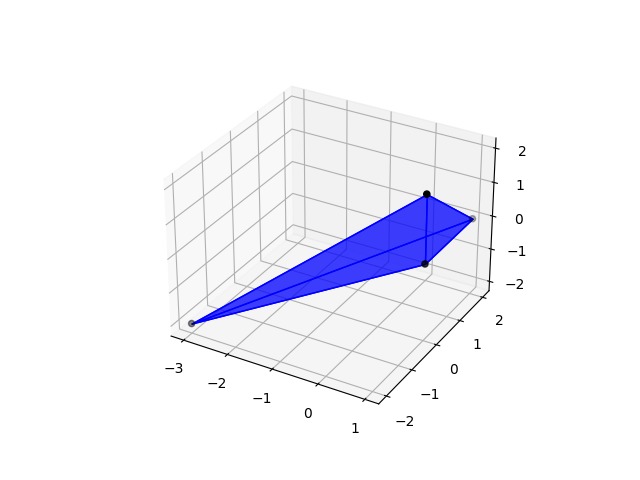}
    \caption{Reflexive polytope from~\eqref{eq:3d_poly_V}.}
    \label{fig:3d_poly}
\end{figure}

\begin{figure}[h]
    \centering
      \begin{subfigure}{0.45\textwidth}
        \includegraphics[width=\textwidth]{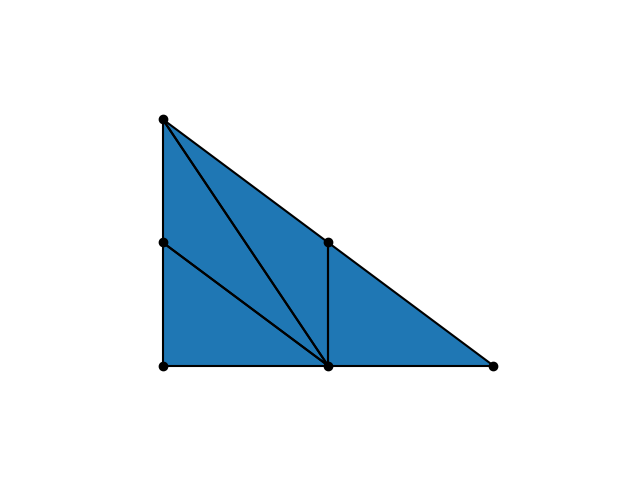}
          \caption{}
      \end{subfigure}
      \hfill
      \begin{subfigure}{0.45\textwidth}
        \includegraphics[width=\textwidth]{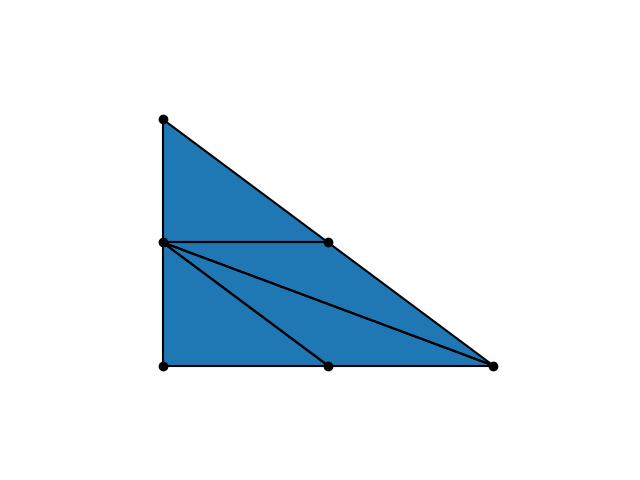}
          \caption{}
      \end{subfigure}
      \begin{subfigure}{0.45\textwidth}
        \includegraphics[width=\textwidth]{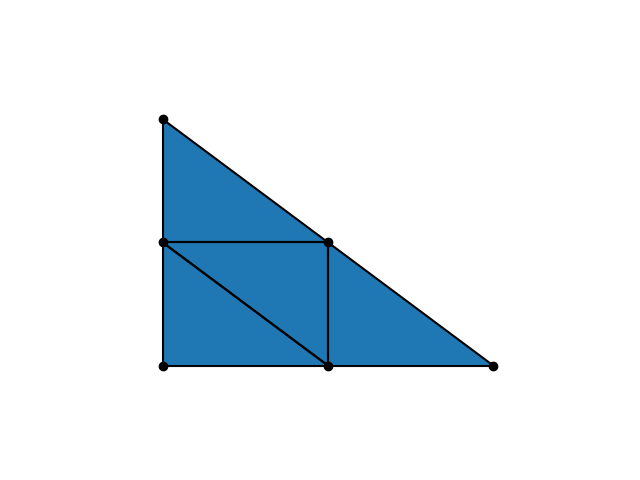}
          \caption{}
          \label{fig:ex_2face_T}
      \end{subfigure}
      \hfill
      \begin{subfigure}{0.45\textwidth}
        \includegraphics[width=\textwidth]{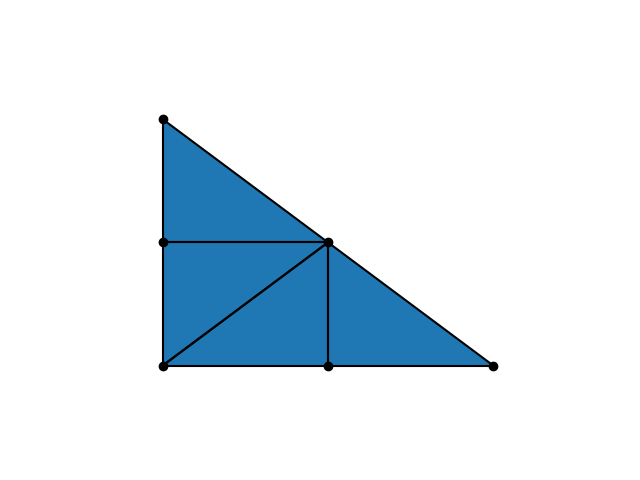}
          \caption{}
      \end{subfigure}
    \caption{All possible fine regular triangulations of a two-face of the polytope $\Delta$ defined by~\eqref{eq:3d_poly_V}.}
    \label{fig:2face_Ts}
\end{figure}

\begin{example}
    Let $\Delta_{1}$ be the three-dimensional reflexive polytope defined as
    \begin{equation}
    \label{eq:3d_poly_V}
        \Delta_{1} = \mathrm{Conv}
        \begin{pmatrix}
            1 && 1 && 1 & -3 \\
            0 && 2 && 0 & -2 \\
            0 && 0 && 2 & -2
        \end{pmatrix} \,,
    \end{equation}
    where Conv denotes the convex hull of the vertices (each vertex a column), as shown in Figure~\ref{fig:3d_poly}.
    There are four two-faces of $\Delta_{1}$, and each of these two-faces has four fine, regular triangulations. Figure~\ref{fig:2face_Ts} shows the possible triangulations for one of the two-faces.
    Therefore, the two-face triangulation states $s\in\mathcal{S}$ will be represented as a $4\times4$ bitlist, \textit{e.g.},
    \begin{equation}
        s = [[1,0,0,0],[1,0,0,0],[0,0,1,0],[0,0,0,1]] \,,
    \end{equation}
    where, for example the sublist $[0,0,1,0]$ corresponds to the two-face triangulation represented in \ref{fig:ex_2face_T}.
\end{example}

With the state space just described, we define the action space $\mathcal{A}$ as all possible changes of two-face triangulations: $\mathcal{T}^{i}_{j}\rightarrow\mathcal{T}^{i}_{k}\ \forall\ i,j,k$.\footnote{
This includes the identity maps $\mathcal{T}^{i}_{j}\rightarrow\mathcal{T}^{i}_{j}$ since the action space must be fixed.}

Finally, we wish to define the fitness function $f:\mathcal{S}\rightarrow[0,1]$ that determines how close a state $s$ is from defining an FRST of the polytope.
To do so, we first combine the two-face triangulations of the state $s\in\mathcal{S}$ to get the full triangulation $\mathcal{T}$ of the polytope.
We then check the fine, regular, and star conditions of $\mathcal{T}$, introduced in Appendix \ref{sec:frst}.
In doing so, we only need for $\mathcal{T}$ to be fine with respect to two-faces since the anticanonical divisor defining the CY does not intersect the divisors associated with points interior to facets.
In our encoding, all the two-face triangulations $\mathcal{T}^{i}_{j}$ are defined to be fine and therefore $\mathcal{T}$ will always be fine with respect to the two-faces and we do not need to check this condition.
Furthermore, any non-star, regular triangulation can be converted into a star triangulation by lowering the height of the origin until it appears as a vertex in all simplices.
Therefore we also do not need to check the star condition, and it remains only to check regularity of the combined triangulation $\mathcal{T}$. 

Recall that for a point configuration $\textbf{A}=(\textbf{a}_{1},...,\textbf{a}_{n})$, the secondary cone $C$ of a regular triangulation $\mathcal{T}$ of $\textbf{A}$ is the space of all height vectors $h=(h_{1},...,h_{n})$ that produce $\mathcal{T}$.
It follows that $\mathcal{T}$ is regular if and only if $C$ is solid, \textit{i.e.}, full-dimensional. 
Now, let $\textbf{A}_{1},...,\textbf{A}_{n}$ be multiple point configurations, with corresponding regular triangulations $\mathcal{T}_{1},...,\mathcal{T}_{n}$ and let $\textbf{A}=\cup_{i=1}^{n}\textbf{A}_{i}$. 
If we embed the secondary cones $\mathcal{C}_{i}$ of $\mathcal{T}_{i}$ into the height space of $\textbf{A}$ and take the intersection $C=\cap_{i=1}^{n}C$, it follows that if $C$ is solid then there exists a regular triangulation $\mathcal{T}$ of $\textbf{A}$ that when restricted to $\textbf{A}_{1},...,\textbf{A}_{n}$ produces the regular triangulations $\mathcal{T}_{1},...,\mathcal{T}_{n}$. 
From this we can determine whether a two-face triangulation state $s$ produces a regular triangulation of the ambient polytope. Firstly, we compute all the secondary cones $C_{i}$ of the two-face triangulations and then compute the intersection $C=\cap_{i=1}^{n}C_{i}$.
The fitness function of a state $s\in\mathcal{S}$ is then defined as:
\begin{equation}
\label{eq:RL_fitness}
    f_{\mathcal{T}}(s) = \frac{N_{p}(F_{\Delta})-\dim{C}}{N_{p}(F_{\Delta})}\,,
\end{equation}
where $ N_{p}(F_{\Delta})$ is the number of points (not interior to two-faces) of the polytope $\Delta$ and therefore this is also the ambient dimension of the intersection cone $C$. 
Given a fitness function $f$ the reward function of a state-action pair is defined as
\begin{equation}
    R(s,a) = f(a(s))-f(s).
\end{equation}
  
\section{Results}\label{sec:results}

To showcase the capability of our RL model at generating datasets of non two-face equivalent (NTFE) FRSTs for a given reflexive polytope, we start with an example. Let $\Delta_{2}$ be the four-dimensional reflexive polytope defined as:
\begin{equation}
    \Delta_{2} = \mathrm{Conv}
    \begin{pmatrix}
        0 & 4 & -2 && 0 && 0 & -2 && 1 \\
        0 & -2 & 2 && 0 && 1 & -2 && 0 \\
        1 & -1 & -1 && 0 && 0 & 2 && 0 \\
        0 & -2 & 0 && 1 && 0 & 1 && 0 
    \end{pmatrix}.
\end{equation}
$\Delta_{2}$ has $17$ points, $19$ two-faces and the maximum number of fine regular triangulations of a two-face is $5$. 
The Hodge numbers for the associated CYs are $h^{1,1}=h^{1,2}=15$.

\begin{figure}[h]
    \centering
    \includegraphics[width=0.6\textwidth]{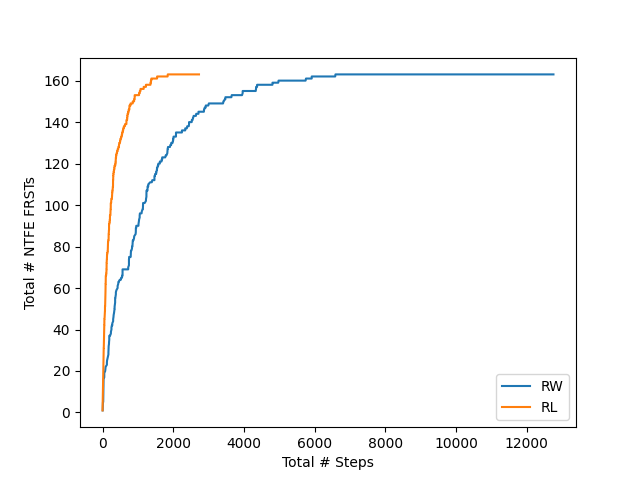}
    \caption{Plot of the total number of non two-face equivalent fine regular star triangulations found against the number of steps taken for a random walk search (RW) and using the trained reinforcement learning model (RL).}
    \label{fig:RL_plot}
\end{figure}

Existing algorithms for finding triangulations, such as the \texttt{triangulations} function in SageMath~\cite{sagemath} and the \texttt{all\_triangulations} function in CYTools~\cite{Demirtas_CYTools_A_Software}, which both rely on TOPCOM~\cite{topcom}, either take too long to run or run out of memory and fail on large polytopes, such as this one. 
We trained our RL algorithm over $1000$ episodes and then used the trained model to generate a dataset of NTFE FRSTs for $\Delta$, terminating when no new triangulations are found for 1000 episodes. For comparison, we also run a random walk search.
The total number of NTFE FRSTs found against total cumulative number of steps taken in both cases are shown in Figure~\ref{fig:RL_plot}.
We see that both searches reach the same total, highlighting that the RL does not introduce further further bias towards generating certain FRSTs, and, moreover, RL is far more efficient and takes far fewer steps.

\subsection{Low $h^{1,1}$}

In a recent paper~\cite{Gendler:2023ujl}, the authors classified the number of equivalence classes of CY threefolds, constructed from four-dimensional reflexive polytopes with $h^{1,1}\in\{1,2,3,4,5\}$ and their triangulations. The authors determined the number of CY classes by, first enumerating all NTFE FRSTs for all polytopes at a given $h^{1,1}$ value, and then computing and comparing Hodge numbers, triple intersection numbers, and second Chern classes. 
The counts from~\cite{Gendler:2023ujl} are presented in Table~\ref{tab:FRST_counts}. 

To test the capability of our RL algorithm at generating complete datasets of NTFE FRSTs for reflexive polytopes we trained the model separately for each polytope with $h^{1,1}\in\{1,2,3,4,5\}$ and used the trained model to search for FRSTs. We run the search until no new triangulations are found in $1000$ episodes, with both methods generating the complete list of NTFE FRSTs are found for every polytope. The total cumulative number of steps taken by the RL model is given in Table~\ref{tab:FRST_counts} as well as the number of steps taken by a random walk search.
These results show that RL can successfully generate complete datasets of NTFE FRSTs and is also more efficient that random walk searches, particularly for polytopes with larger $h^{1,1}$. 

\begin{table}[h!]
\centering
\addtolength{\leftskip}{-1.75cm}
\addtolength{\rightskip}{-1.75cm}
 \begin{tabular}{|| c c c c c c c||} 
 \hline
 $h^{1,1}$ & \# polys & \# FRSTs & \# NTFE FRSTs & \# FRST classes & RW \# Steps & RL \# Steps \\ [0.5ex] 
 \hline\hline
 $1$ & $5$ & $5$ & $5$ & $5$ & $0$ & $0$ \\ 
 $2$ & $36$ & $48$ & $39$ & $36$ & $7$ & $6$ \\
 $3$ & $244$ & $526$ & $306$ & $275$ & $144$ & $107$ \\
 $4$ & $1197$ & $5348$ & $2022$ & $1774$ & $2658$ & $2292$ \\
 $5$ & $4990$ & $57050$ & $13814$ & $11847$ & $41902$ & $33148$ \\ [1ex] 
 \hline
 \end{tabular}
 \caption{Counts of non two-face equivalent (NTFE) fine regular star triangulations (FRSTs) of reflexive polytopes with low $h^{1,1}$, with the number of steps taken by a random walk (RW) search and by the trained reinforcement learning (RL) model to generate them all. Also listed in the third and fifth columns are the total numbers of FRSTs and FRST classes, respectively, at each $h^{1,1}$ value.}
\label{tab:FRST_counts}
\end{table}

\subsection{Bundles}\label{sec:bundles}

Analogous to the targeted search presented in~\cite{Berglund:2023ztk} for reflexive polytopes with GAs, we now present an example of a targeted search for FRSTs of reflexive polytopes by modifying our RL model. This search is inspired by~\cite{Abel:2023zwg}, and complements other machine learning approaches to CY bundles~\cite{Klaewer:2018sfl,Brodie:2019dfx,Deen:2020dlf}.
In ~\cite{Abel:2023zwg}, the authors consider $E_{8}\times E_{8}$ heterotic string compactified on smooth CY threefolds $X$ with holomorphic vector bundles $V$. They showcase how GAs can be used, given a CY threefold $X$, to find suitable bundles $V$ that satisfy a set of conditions, namely i) anomaly cancellation, ii) slope-stability, and iii) appropriate particle spectrum. They focus on the particular case where $V$ is a sum of line bundles, in which case the conditions are straightforward to check. 
Specifically, $V$ is taken to be a rank-$5$ line bundle sum $V=\oplus_{a=1}^{5}L_{a}$, where $L_{a}=\mathcal{O}_{X}(L_{a})$, so that the resulting model has $SU(5)\times S(U(1)^{5})$ symmetry. The line bundle $L_{a}$ has first Chern class $c_{1}(L_{a})=k_{a}^{i}J_{i}$, where $k_{a}\in\mathbb{Z}^{h}$ are integer vectors and $(J_{1},...,J_{h})$ is a suitably chosen basis of $H^{2}(X,\mathbb{Z})$, where $h=h^{1,1}(X)$. Therefore, the line bundle $L_{a}$ can be written as $\mathcal{O}_X(k_{a}^{1}, k_{a}^{2}, \ldots, k_{a}^{h})$. In general, $k_{a}^{I}$ are a mixture of positive and non-positive integers. The five integer vectors $(k_{1},...,k_{5})$ therefore uniquely specify the line bundle sum $V$. 

In our targeted RL search we focus on the first three conditions (C1)--(C3) from~\cite{Abel:2023zwg}:
\begin{enumerate}
    \item \textbf{$E_{8}$ Embedding}: 
    \begin{equation}
        c_{1}(V) = \sum_{a=1}^{5}k_{a} = 0 \,.
    \end{equation}
    This condition ensures that the structure group of $V$ is $S(U(1)^{5})$ and not a smaller subgroup. 
    \item \textbf{Anomaly Cancellation}:
    \begin{equation}
        c_{2,i}(V) = -\frac{1}{2}d_{ijk}\sum_{a=1}^{5}k_{a}^{j}k_{a}^{k} \leq c_{2,i}(TX) \,,
    \end{equation}
    for all $i=1,...,h$, where $d_{ijk}$ denotes the triple intersection numbers and $c_{2}(TX)$ denotes the second Chern class of the tangent bundle of $X$ relative to the basis $(J_{1},...,J_{h})$.
    \item \textbf{Poly-Stability}: There exists a non-trivial solution $t^{i}$ to 
    \begin{equation}
        \mu(L_{a}) = d_{ijk}k_{a}^{i}t^{j}t^{k} \,,
    \end{equation}
    for all $a=1,...,5$, such that $J=t^{i}J_{i}$ is in the interior of the Kähler cone, which in the case where $V=\oplus_{a=1}^{5}L_{a}$ corresponds to $t^{i}>0$. This check is computationally expensive but can be replaced with the weaker condition that the matrices $M_{a}=(d_{ijk}k_{a}^{i})$ for $a=1,...,5$, and any linear combination $v^{a}M_{a}$, all have at least one negative and one positive entry. Practically, considering $v_{a}$, with entries in $\{-2,-1,0,1,2\}$ provides a strong enough check.
\end{enumerate}

In contrast to the GA search performed in~\cite{Abel:2023zwg}, where the CY geometry was fixed, we search the space of CYs\footnote{The space of CYs is restricted to those originating from the same polytope but with different FRSTs.} and line bundles simultaneously. We extend the state encoding from before to include the matrix $(k_{1},...,k_{5})$ that specifies the line bundle sum. 
Furthermore, we modify the fitness function~\eqref{eq:RL_fitness} to include terms that penalise individuals that do not satisfy the constraints (C1)--(C3):
\begin{equation}
    f_{\mathcal{T}-V}(s) = w_{1}f_{\mathcal{T}}(s) + w_{2}f_{\text{GUT}} + w_{3}f_{\text{anom}} + w_{4}f_{\text{slope}} \,,
\end{equation}
where $f_{\text{GUT}}$, $f_{\text{anom}}$, $f_{\text{slope}}$ are the contributions associated with the $E_{8}$ embedding, cancellation of anomalies, and slope stability, respectively, and the $w_{i}\in\mathbb{R}^{>0}$ are weights, which we set to $w_{1}=w_{2}=w_{3}=w_{4}=\frac{1}{4}$. 
The $E_{8}$ GUT embedding, anomaly cancellation, and slope stability condition contributions are given explicitly as
\begin{equation}
    f_{\text{GUT}} = \frac{1}{h}\sum_{i=1}^{h} 
    \begin{cases}
        1 & \text{if\quad $\sum_{a=1}^{5}k_{a}^{i}=0$} \,, \\
        0 & \text{if\quad $\sum_{a=1}^{5}k_{a}^{i}\neq0$} \,,
    \end{cases}
\end{equation}
\begin{equation}
    f_{\text{anom}} = \frac{1}{h}\sum_{i=1}^{h}
    \begin{cases}
        1 & \text{if \quad $-\frac{1}{2}\sum_{a=1}^{5}\sum_{j=1}^{h}\sum_{k=1}^{h} d_{ijk} k_{a}^{j} k_{a}^{k} \leq c_{2}$} \,, \\
        0 & \text{if \quad $-\frac{1}{2}\sum_{a=1}^{5}\sum_{j=1}^{h}\sum_{k=1}^{h} d_{ijk} k_{a}^{j} k_{a}^{k} > c_{2}$} \,,
    \end{cases}
\end{equation}
\begin{equation}
    f_{\text{slope}} = \frac{1}{3125}\sum_{n=1}^{3125}
    \begin{cases}
        1 & \text{if $\max{(v_{n})^{a}M_{a}}>0$ and $\min{(v_{n})^{a}M_{a}}<0$} \,, \\
        0 & \text{if not $\max{(v_{n})^{a}M_{a}}>0$ and $\min{(v_{n})^{a}M_{a}}<0$} \,,
    \end{cases}
\end{equation}
where $M_{a} = d_{ijk} k_{a}^{i}$ and $v_{n} \in \mathbb{Z}^{5}$ with entries in $\{-2,-1,0,1,2\}$. There are $5^5=3125$ $v_{n}$ vectors which explains this value appearing in the $f_{\text{slope}}$ formula. The $\frac{1}{h}$ and $\frac{1}{3125}$ factors ensure that the maximum value for each contribution is $1$ and therefore each condition contributes equally to the fitness. 

In~\cite{He:2013ofa}, the authors determine all CY threefolds constructed from the KS database, which have non-trivial first fundamental group, and classify all $SU(5)$ and $SO(10)$ line bundle models on these manifolds which satisfy the conditions (C1)--(C3). We extract one of these example manifolds from~\cite{He:2013ofa}, which we know admit appropriate $SU(5)$ bundles and apply our RL model to look for compatible $(\mathcal{T},k)$ triangulation-bundle pairs. The example we consider is defined by\footnote{In~\cite{He:2013ofa} this CY is labeled as $X_{13}$.}
\begin{equation}
\label{eq:4d_poly_V}
    \Delta_{3} = \mathrm{Conv}
    \begin{pmatrix}
        0 & 0 && 0 & -2 & 0 && 0 && 2 & 0 \\
        -1 & -1 && 1 & 0 & 0 && 0 && 0 & -1 \\
        -1 & 1 && 0 & 0 & -1 && 1 && 0 & 0 \\
        -1 & 1 && 0 & -1 & 0 && 0 && 1 & 0 
    \end{pmatrix} \,.
\end{equation}

From~\cite{He:2013ofa}, we know that all appropriate line bundle sums of these manifolds have $k$ matrices with integers in the range $[-4,4]$. Therefore, we use this range to define our state space and train the RL model over 1000 episodes. An episode ends either when a $(\mathcal{T},k)$ pair is found, such that $\mathcal{T}$ is an FRST of $\Delta$ and $k$ defines a line bundle sum that satisfies all the constraints, or the number of steps taken reaches $1000$ and no terminal state is found. 
Once the model has been trained, we use it to search for  $(\mathcal{T},k)$ pairs. 
An example of a triangulation and line bundle sum pair found by the RL after training is given below.

\begin{example}
    An FRST of the reflexive polytope defined by~\eqref{eq:4d_poly_V} with compatible line bundle sum found by the trained RL model is given by:
    \begin{equation}
        \begin{split}
            \mathcal{T} &= \{\{0,1,3,4,5\},\{0,1,3,4,6\},\{0,1,3,5,7\},\{0,1,3,6,7\},\{0,1,4,5,8\}, \\
            & \{0,1,4,6,8\},\{0,1,5,7,8\},\{0,1,6,7,8\},\{0,2,3,4,5\},\{0,2,3,4,6\}, \\
            & \{0,2,3,5,7\},\{0,2,3,6,7\},\{0,2,4,5,8\},\{0,2,4,6,8\},\{0,2,5,7,8\},\{0,2,6,7,8\}\}
        \end{split}
    \end{equation}
    \begin{equation}
        k = 
        \begin{pmatrix}
            1 & -2 & -1 & 1 & 1 \\
            0 & 1 & -1 & 0 & 0 \\
            -1 & 0 & 3 & -1 & -1 \\
            0 & -1 & 1 & 0 & 0 
        \end{pmatrix} \,,
    \end{equation}
    where in $\mathcal{T}$ the $0$ index corresponds to the origin and the indices $1-8$ correspond to the points whose coordinates are given in columns $1-8$ of~\eqref{eq:4d_poly_V}.
\end{example}

\subsection{Fibrations}\label{sec:fibrations}

Recently there has been progress in identifying the existence of certain fibration structures of a toric CY using the geometric data of the polytope~\cite{Huang:2018vup, Huang:2019pne, Rohsiepe:2005qg, Knapp:2011ip}. Explicit construction of the fibration structure involves finding a reflexive subpolytope $\Delta'\subset \Delta$ and a triangulation compatible with $\Delta'$ and $\Delta$. A brief overview of fibration structures is given in Appendix~\ref{appendix:fibrations}. From the results of~\cite{Huang:2019pne}, existence of a two-dimensional reflexive subpolytope inside a four-dimensional reflexive polytope implies the existence of compatible triangulations yielding an elliptic-fibration structure in the corresponding family of CY threefolds. However, this in general is not the case. The only other exception being $\mathrm{codim}=1$ subpolytopes, where there always exists compatible triangulations.
In this section, we apply the same RL model used in the previous section to search for reflexive subpolytopes and compatible triangulations.

\subsubsection{Subpolytopes}

\begin{figure*}[htb]
    \centering
    \includegraphics[width=0.5\textwidth]{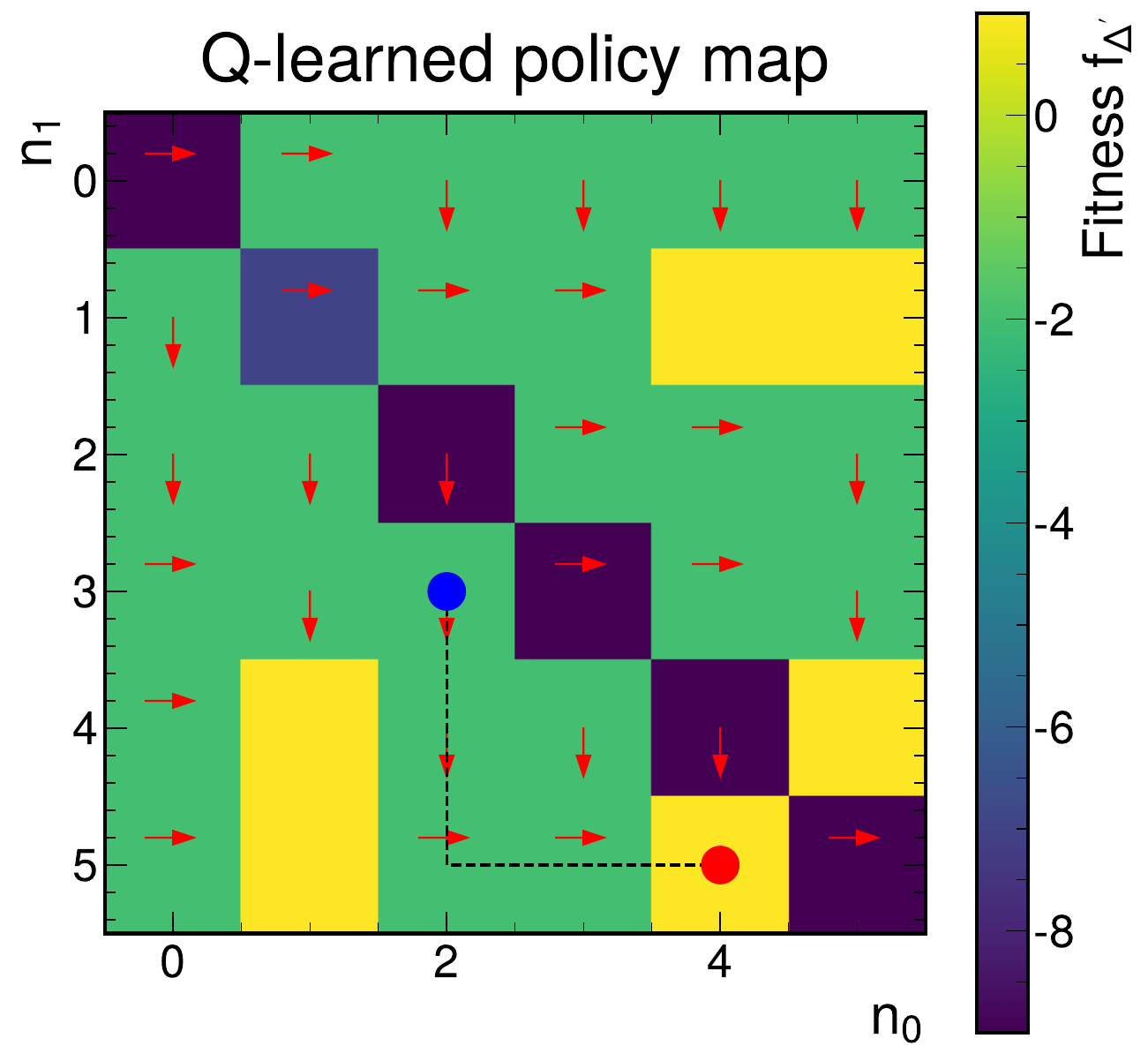}
    \caption{The fitness and Q-learned policy maps on the state space of~\eqref{eq:policyMapExamplePolytope}. The policies are denoted by the arrows $\{{\color{red}\rightarrow},{\color{red}\downarrow}\}$. The terminal point of the path, starting at the {\color{blue} blue dot} and terminating at the {\color{red} red dot}, corresponds to the elliptic fiber~\eqref{eq:policyMapExampleFiber}. The coordinates $(n_0, n_1)$ correspond to vertices $(v_{n_0}, v_{n_1})$ spanning the plane of $\Delta'$.}\label{fig:rewardAndPolicy}
\end{figure*}

We begin by applying our RL model to search for reflexive subpolytopes. The encoding of the subpolytopes $\Delta'$ is outlined in Appendix~\ref{appendix:fibrations}. An algorithm which uses this encoding to find fibration structures of toric CYs, by enumerating all possible subspaces, has been proposed in~\cite{Rohsiepe:2005qg}. We instead develop a Q-learning based method to search for such structures. We define the fitness function by penalizing states that do not meet the required subpolytope conditions $\Delta'$. More precisely, we have:
\begin{equation}
\label{eq:rewardVirtFibrations}
	f_{\Delta'}(s) = f_{\text{reflex}}(s) + f_{\text{rank}}(s) \,,
\end{equation}
where the two terms check whether the subpolytope is reflexive and has the appropriate dimension. Explicitly,
\begin{align}
	f_{\text{reflex}}(s)&:= 
    \begin{cases}
 	  1\quad\text{if}~\Delta'~\text{is reflexive} \,, \\
 	  -1\quad\text{otherwise} \,,
    \end{cases} \\
    f_{\text{rank}}(s)&:= -(\dim \Delta' - d)^2 \,,
\end{align}
where $\Delta'$ is the subpolytope associated to the state $s$ and $d$ is the desired subpolytope dimension.
This provides a ``maze''-like environment for Q-learning. As an example, consider the four-dimensional reflexive polytope:
\begin{equation}
\label{eq:policyMapExamplePolytope}
	\Delta_{4} = \mathrm{Conv}
    \begin{pmatrix}
        1 && 0 && 0 && 0 & -1 & -1 \\
        0 && 1 && 0 && 0 & -1 & -1 \\
        0 && 0 && 1 && 0 & 0 & -1 \\
        0 && 0 && 0 && 1 & 0 & -1
    \end{pmatrix}\,.
\end{equation} 
The space of states $\mathcal{S}$ corresponding to virtual two-fibrations is two-dimensional. We plot the fitness $f_{\Delta'}$ and Q-learned policy maps in Figure~\ref{fig:rewardAndPolicy}.
 The polytope~\eqref{eq:policyMapExamplePolytope} admits only a single virtual two-fibration structure given by the subpolytope:
 \begin{equation}
 \label{eq:policyMapExampleFiber}
     \Delta_{4}' = \mathrm{Conv}
     \begin{pmatrix}
        1 && 0 & -1 \\
        0 && 1 & -1 \\
        0 && 0 & 0 \\
        0 && 0 & 0 
     \end{pmatrix} \,.
 \end{equation}
 
To demonstrate the generality of this approach, we consider the case when $\Delta$ is a five-dimensional reflexive polytope:

\begin{example} 
Let $\Delta_{5} \subset N\simeq \mathbb{Z}^5$ be a reflexive polytope:
\begin{equation}
    \Delta_{5} = \mathrm{Conv}
    \begin{pmatrix}
        -2 & 0 & 0 & 1 & 1 & 2 \\
        0 & -1 & 0 & 0 & 0 & 1 \\
        -1 & -1 & 1 & 0 & 0 & 1 \\
        2 & 1 & 2 & -3 & -1 & -5 \\
        5 & 0 & -1 & -3 & -3 & -2
    \end{pmatrix}\,.
\end{equation}
The Q-learned two-dimensional reflexive subpolytopes are generated by vertex matrices:
\begin{gather}
    \begin{pmatrix}
       0 & -2 & 0 & -1 & 1 \\
       0 & 0 & 0 & 0 & 0 \\
       0 & -1 & 1 & 0 & 0 \\
       0 & 2 & 2 & 2 & -2 \\
       0 & 5 & -1 & 2 & -2
    \end{pmatrix},\qquad \begin{pmatrix}
    0 & -1 & 1 & 1 & 1 \\
    0 & 0 & 0 & 0 & 0 \\
    0 & 0 & 0 & 0 & 0 \\
    0 & 2 & -2 & -2 & -2 \\
    0 & 2 & -3 & -1 & -2
    \end{pmatrix}\,,
\end{gather}
while the Q-learned three-dimensional reflexive subpolytopes are generated by:
\begin{gather}
\begin{pmatrix}
    0 & -2 & 0 & -1 & 1 & 1 & 1 \\
    0 & 0 & 0 & 0 & 0 & 0 & 0 \\
    0 & -1 & 1 & 0 & 0 & 0 & 0 \\
    0 & 2 & 2 & 2 & -2 & -2 & -2 \\
    0 & 5 & -1 & 2 & -3 & -1 & -2 
\end{pmatrix}\,,\quad
\begin{pmatrix}
    0 & 0 & 2 & -1 & 1 & 1 & 1 \\
    0 & -1 & 1 & 0 & 0 & 0 & 0 \\
    0 & -1 & 1 & 0 & 0 & 0 & 0 \\
    0 & 1 & -5 & 2 & -2 & -2 & -2 \\
    0 & 0 & -2 & 2 & -3 & -1 & -2
\end{pmatrix}\, \\
\begin{pmatrix}
    0 & 1 & 1 & -1 & 1 & 1 & 1 \\
    0 & 0 & 0 & 0 & 0 & 0 & 0 \\
    0 & 0 & 0 & 0 & 0 & 0 & 0 \\
    0 & -3 & -1 & 2 & -2 & -2 & -2 \\
    0 & -3 & -3 & 2 & -3 & -1 & -2
\end{pmatrix}\,.
\end{gather}
Finally, the Q-learned four-dimensional reflexive subpolytopes are generated by:
\begin{gather}
    \begin{pmatrix}
        0 & -2 & 0 & 0 & 2 & -1 & 1 & 1 & 1 \\
        0 & 0 & -1 & 0 & 1 & 0 & 0 & 0 & 0 \\
        0 & -1 & -1 & 1 & 1 & 0 & 0 & 0 & 0 \\
        0 & 2 & 1 & 2 & -5 & 2 & -2 & -2 & -2 \\
        0 & 5 & 0 & -1 & -2 & 2 & -3 & -1 & -2 
    \end{pmatrix}\,,\quad
    \begin{pmatrix}
        0 & -2 & 0 & 1 & 1 & -1 & 1 & 1 & 1 \\
        0 & 0 & 0 & 0 & 0 & 0 & 0 & 0 & 0 \\
        0 & -1 & 1 & 0 & 0 & 0 & 0 & 0 & 0 \\
        0 & 2 & 2 & -3 & -1 & 2 & -2 & -2 & -2 \\
        0 & 5 & -1 & -3 & -3 & 2 & -3 & -1 & -2 
    \end{pmatrix}\, \\
    \begin{pmatrix}
        0 & 0 & 1 & 1 & 2 & -1 & 1 & 1 & 1 \\
        0 & -1 & 0 & 0 & 1 & 0 & 0 & 0 & 0 \\
        0 & -1 & 0 & 0 & 1 & 0 & 0 & 0 & 0 \\
        0 & 1 & -3 & -1 & -5 & 2 & -2 & -2 & -2 \\
        0 & 0 & -3 & -3 & -2 & 2 & -3 & -1 & -2 
    \end{pmatrix}\,.
\end{gather}
The four-dimensional virtual fibrations corresponding to the reflexive subpolytopes above are guaranteed to have a compatible set of fans by the results of \cite{Rohsiepe:2005qg}.
\end{example}

\subsubsection{Compatible triangulations}

In order to incorporate triangulations into the search, we extend the state space. 
In particular, the state space for searching for $d$-fibration structures together with compatible triangulations, is given as a product: $\mathcal{S} = \mathcal{S}_{\mathrm{triang}.} \times \mathcal{S}_{\mathrm{fib}.}$, where $\mathcal{S}_{\mathrm{triang}.}$ corresponds to the state space of triangulations defined in Section~\ref{sec:model}, and $\mathcal{S}_{\mathrm{fib}.}$ corresponds to the state space of virtual fibrations defined in Appendix \ref{appendix:fibrations}. The action space $\mathcal{A}$ is constructed in the same manner. The fitness function is extended by adding a penalty ensuring compatibility of the fan corresponding to the triangulation with the virtual fibration structure:
\begin{equation}\label{eq:rewardFibrations}
    f_{\Delta'-\mathcal{T}}(s_{1},s_{2}) = f_{\Delta'}(s_{1}) +  f_{\mathcal{T}}(s_{2}) + f_{\text{compat}}(s_{1},s_{2}),
\end{equation}
where $s_{1},s_{2}$ describe the subpolytope and two-face triangulations respectively, and $f_{\text{compat}}$ is the contribution associated to the compatibility condition. In particular,\footnote{Note that it might be the case that even if there exists a fibration structure, there is no compatible triangulation generated using two-face encoding. In those cases, for each state we compute neighboring triangulations using bistellar flips \cite{cytools}.} \cite{Knapp:2011ip}.
\begin{equation}
	f_{\text{compat}}(s_{1},s_{2}):= 
    \begin{cases}
 	  1\quad\text{if the restriction}~\mathcal{T}\vert_{\Delta'}~\text{is an FRST of}~\Delta' \,, \\
 	  0\quad\text{otherwise} \,.
    \end{cases}
\end{equation}
As a proof-of-concept, an example using the reward function \eqref{eq:rewardFibrations} is presented below.

\begin{example} 
Let $\Delta_{6}\subset N \simeq\mathbb{Z}^4$ denote the reflexive polytope:
\begin{gather}\label{eq:fibrationEg1}
    \Delta_{6} = \mathrm{Conv}
    \begin{pmatrix}
        -1 && 0 && 0 && 1 & -1 & 0 && 0 \\
        -1 && 0 && 1 && 0 & 0 & -1 && 0 \\
        1 && 0 && 0 && 0 & -1 & -1 && 1 \\
        -1 && 1 && 0 && 0 & 0 & 0 && 0
    \end{pmatrix}\,.
\end{gather}
Two-fibration structures obtained using the trained Q-learning model, given by reflexive two-dimensional subpolytopes $\Delta'$ and $\Delta''$ and their respective compatible triangulations are shown in Figure~\ref{fig:fibrationEg1}.
The vertices of $\Delta'$ and $\Delta''$ are:
\begin{gather}\label{eq:fibrationEg1F1}
    \Delta' = \mathrm{Conv}
    \begin{pmatrix}
        0 & -1 && 0 && 1 \\
        0 & 0 && 0 && 0 \\
        0 & -1 && 1 && 0 \\
        0 & 0 && 0 && 0 
    \end{pmatrix}\,,\qquad \Delta'' = \mathrm{Conv}
    \begin{pmatrix}
       0 & 0 && 0 && 0 \\
       0 & -1 && 0 && 1 \\
       0 & -1 && 1 && 0 \\
       0 & 0 && 0 && 0
    \end{pmatrix}\,.
\end{gather}
It is easy to see that both $\Delta'$ and $\Delta''$ are indeed reflexive.
We may also look for three-fibration structures of \eqref{eq:fibrationEg1} (corresponding to K3 fibered CYs). In particular, $\Delta$ admits a three-dimensional reflexive subpolytope $\Delta{\mathrm{K3}}$ given by:
\begin{gather}
    \Delta{\mathrm{K3}} = \mathrm{Conv}
    \begin{pmatrix}
         0 && 0 && 1 & -1 & 0 && 0 \\
         0 && 1 && 0 & 0 & -1 && 0 \\
         0 && 0 && 0 & -1 & -1 && 1 \\
         0 && 0 && 0 & 0 & 0 && 0 
    \end{pmatrix}\,,
\end{gather}
the visualization of a triangulation compatible with the virtual three-fibration $\Delta{\mathrm{K3}}\hookrightarrow \Delta$ is shown in Figure~\ref{fig:fibrationEg1K3}.
\begin{figure}
    \centering
    \begin{subfigure}[b]{0.5\textwidth}
        \centering
        \begin{tikzpicture}[x=1.0\textwidth,y=0.25\textheight]
            \clip (-0.5,-0.5) rectangle (0.5,0.4);
            \node[inner sep=0pt,anchor=center] (0,0) {\includegraphics[width=1.0\textwidth,keepaspectratio]{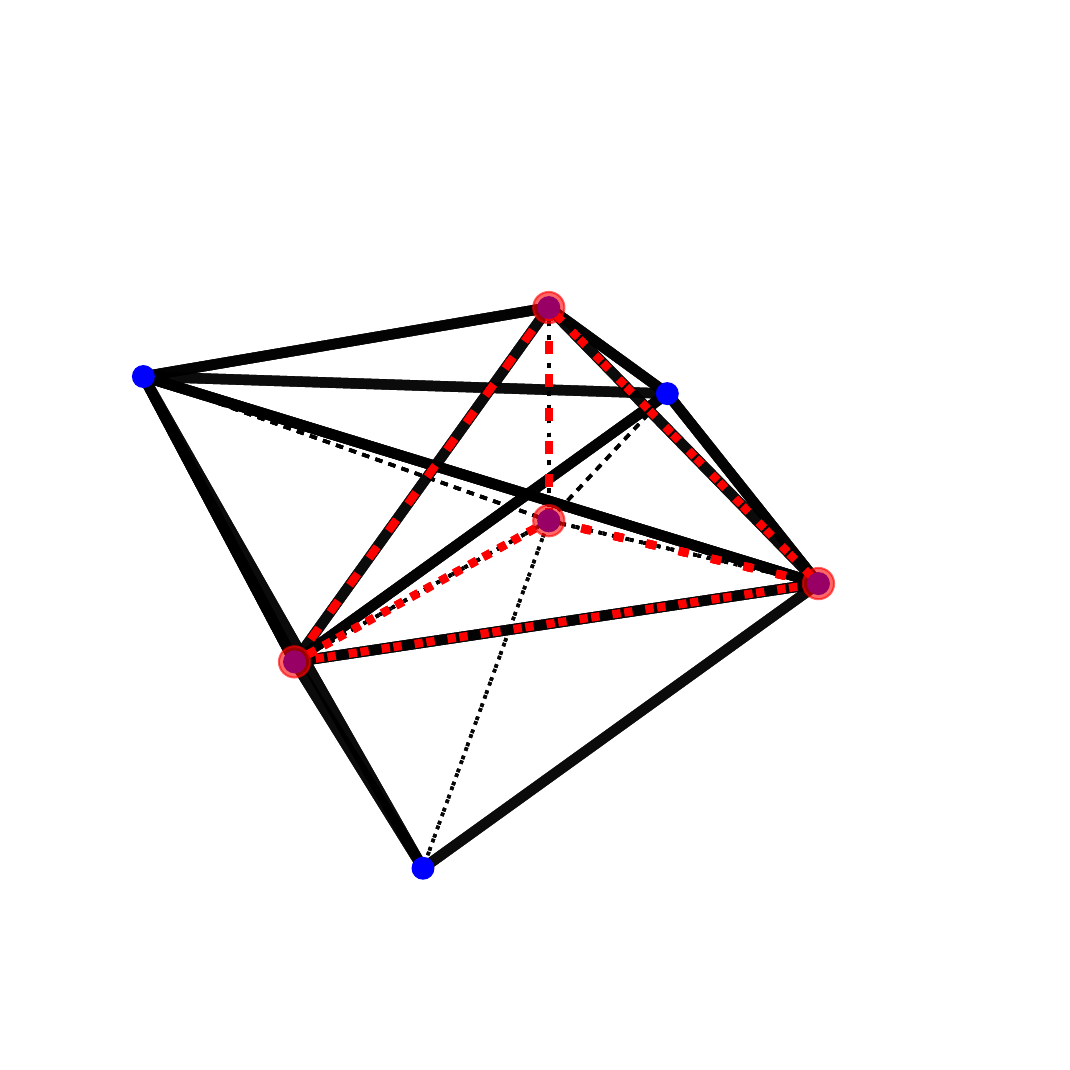}};
        \end{tikzpicture}
        \caption{}
    \end{subfigure}%
    \begin{subfigure}[b]{0.5\textwidth}
        \centering
        \begin{tikzpicture}[x=1.0\textwidth,y=0.25\textheight]
            \clip (-0.5,-0.5) rectangle (0.5,0.4);
            \node[inner sep=0pt,anchor=center] (0,0) {\includegraphics[width=1.0\textwidth,keepaspectratio]{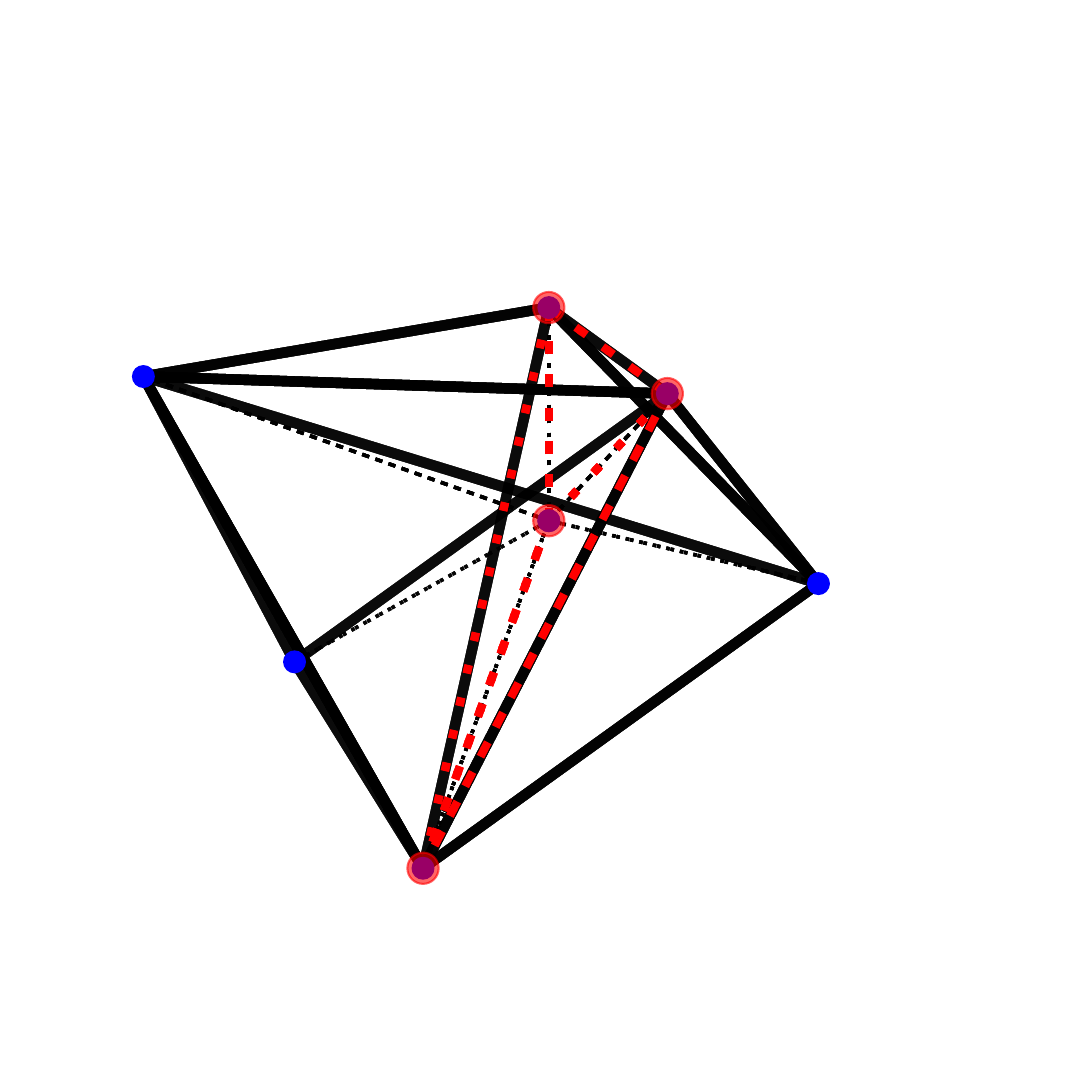}};
        \end{tikzpicture}
        \caption{}
    \end{subfigure}
    \caption{Two-fibrations $\Delta',\Delta''\hookrightarrow \Delta$ with compatible triangulations found using the trained Q-learning model with reward function \eqref{eq:rewardFibrations}. The vertices are projected onto $\mathbb{Z}^3$ by taking first three components in $\mathbb{Z}^4$. The {\color{red} red dots and edges} correspond to the subpolytopes $\Delta'$ (a) and $\Delta''$ (b) \eqref{eq:fibrationEg1F1}.}
    \label{fig:fibrationEg1}
\end{figure}
\begin{figure}[htb]
    \centering

    \begin{tikzpicture}[x=0.6\textwidth,y=0.25\textheight]
        \clip (-0.5,-0.5) rectangle (0.5,0.4);
        \node[inner sep=0pt,anchor=center] (0,0) {\includegraphics[width=0.5\textwidth,keepaspectratio]{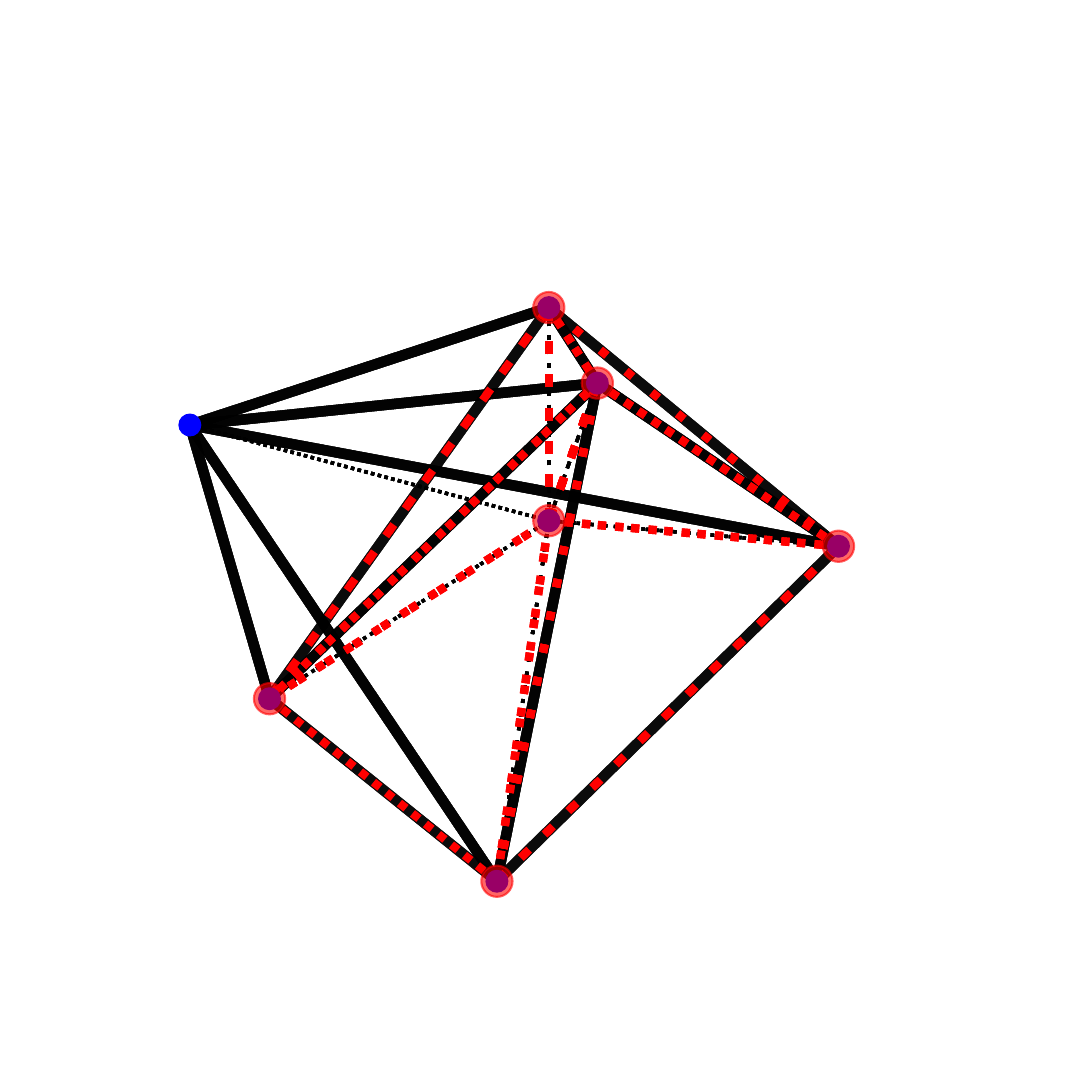}};
    \end{tikzpicture}
    
    \caption{A three-fibration $\Delta{\mathrm{K3}}\hookrightarrow \Delta$ with compatible triangulation computed using Q-learning with reward function \eqref{eq:rewardFibrations}.}\label{fig:fibrationEg1K3}
\end{figure}
\end{example}

\section{Outlook}\label{sec:outlook}

Following on from the work carried out in~\cite{Berglund:2023ztk}, in this paper we have shown how reinforcement learning (RL) can be efficiently used to generate fine regular star triangulations (FRSTs) of four-dimensional reflexive polytopes. Such triangulations provide resolutions of non-terminal singularities in the ambient toric variety, and consequently the Calabi-Yau (CY) threefold hypersurfaces, constructed from the corresponding polytope. 
As detailed in~\cite{demirtas2020bounding}, the total number of FRSTs of four-dimensional reflexive polytopes in the complete classification~\cite{kreuzer2000complete}, is as large as $1.53\times 10^{928}$.\footnote{Triangulations of the largest polytope, with $h^{1,1}=491$, make up the majority of all triangulations. Some of the two-faces of this polytope are too large that it is not feasible to compute all two-face triangulations. It is for this reason that we have have not presented results for generating triangulations of this polytope with RL. It is possible, as was done in \cite{MacFadden:2024him}, to take a sample of triangulations of the two-faces and use these to sample FRSTs of the ambient polytope, however this produces a biased sample.} It is therefore not feasible to generate a complete list of triangulations and instead we would like a way of generating fair samples. In the context of string theory, we would like to generate FRSTs that produce CY threefolds satisfying certain physical constraints. 

Our results show that RL can generate complete datasets of FRSTs for reflexive polytopes with low $h^{1,1}$, where the total number of FRSTs is manageable. Specifically, we regenerated all non two-face equivalent FRSTs with $h^{1,1}\in\{1,2,3,4,5\}$ matching the total found in~\cite{Gendler:2023ujl}. Following this, we considered the particular case of $E_{8}\times E_{8}$ heterotic string compactified on a smooth CY threefolds $X$ with holomorphic vector bundles $V$. Fixing the reflexive polytope and searching the space of FRSTs and line bundle sums $V=\oplus_{a=1}^{5}L_{a}$, where $L_{a}=\mathcal{O}_{X}(L_{a})$, we were able to find examples that satisfy anomaly cancellation and poly-stability constraints. 
We also demonstrated how our RL model can be used look for CY manifolds that admit certain fibration structures. Starting from a reflexive polytope, we were able to find three-dimensional reflexive subpolytopes and compatible FRSTs that together define smooth K3 fibered CYs. 
These targeted search examples, as well as those presented in~\cite{Berglund:2023ztk,MacFadden:2024him}, illustrate how one can design a dedicated search for CY manifolds with prescribed properties as required for the intended string compactification.

\section*{Acknowledgements}
We thank Andre Lukas for collaboration in initial stages of this project.
We are grateful to Nate MacFadden, Andreas Schachner, and Elijah Sheridan for discussions on two-face restrictions and for sharing a pre-arXiv draft of~\cite{MacFadden:2024him}.
We also thank the Pollica Physics Center, Italy, and the organizers of the 2023 Workshop on Machine Learning for a very stimulating environment which is where this project was initiated.
PB and GB are supported in part by the Department of Energy grant DE-SC0020220.
YHH is supported by STFC grant ST/J00037X/2.
E.~Heyes is supported by City, University of London and the States of Jersey.
E.~Hirst is supported by Pierre Andurand.
VJ is supported by the South African Research Chairs Initiative of the Department of Science and Innovation and the National Research Foundation.

\appendix
\section{Toric geometry}\label{sec:toric}

We briefly recall relevant aspects of toric geometry with the goal of understanding fine regular star triangulations of reflexive polytopes and fibration structures of the resulting toric varieties. 

\subsection{Reflexive polytopes}

Given an $n$-dimensional lattice polytope $\Delta\subset M$, one can construct a compact toric variety $X_{\Delta}$ of complex dimension $n$.
In short, one constructs the normal fan $\Sigma_{\Delta}$ as follows: for a face $\theta$ of $\Delta$, let $\sigma_{\theta} \subset N$ be the dual of the cone:
\begin{equation}
    \sigma_{\theta}^{\vee} := \{ \lambda(u-u') | u \in \Delta, u' \in \theta, \lambda \geq 0 \} \subset M \,.
\end{equation}
The normal fan is then given as $\Sigma_{\Delta}:=\{\sigma_{\theta}\}$ for all faces $\theta$ of $\Delta$.
From the normal fan, the construction of the compact projective toric variety $X_{\Delta}$ follows the usual procedure, where each cone $\sigma_{\theta}$ gives rise to an affine toric variety $U_{\sigma_{\theta}}=\text{Spec}(\mathbb{C}[\sigma_{\theta}^{\vee}\cap M])$ and one glues these patches together. 

A polytope is said to satisfy the \textit{IP property} if it has only a single interior lattice point, taken to be the origin.
\begin{definition}
    A lattice polytope $\Delta$ is called \textbf{reflexive} if it satisfies the IP property and if its dual $\Delta^*$ is also a lattice polytope that satisfies the IP property.
    Equivalently, the lattice polytope is IP and all of its facets are a unit distance from the origin.
\end{definition}
The reflexive polytope is only defined up to $GL(n,\mathbb{Z})$ transformations of the coordinates of the vertices; equivalent reflexive polytopes have the same normal form (which is a specific representative from the $GL(n,\mathbb{Z})$ equivalence class).

The connection between CY manifolds and reflexive polytopes is given by the following theorem due to Batyrev~\cite{batyrev1993dual,batyrev1994calabiyau}:
\begin{theorem}
\label{theorem:batyrev}
    Let $\Delta \subset M$ be an $n$-dimensional lattice polytope and $X_{\Delta}$ the corresponding $n$ complex dimensional toric variety.
    If $\Delta$ is reflexive then it follows that $X_{\Delta}$ is Gorenstein Fano with at most canonical singularities and moreover the zero locus of a generic section of the anticanonical bundle $-K_X$ is a CY variety $\mathcal{M}$ of complex dimension $n-1$.
\end{theorem}
The mirror CY $\mathcal{W}$ is similarly obtained from the dual polytope. 

\subsection{Fine regular star triangulations}\label{sec:frst}

The variety $X_{\Delta}$ generated by the reflexive polytope $\Delta$ may be singular.
If it is too singular, then the CY variety $\mathcal{M}$ may not be smooth.
We wish to find a resolution of the singularities given by a birational morphism $\pi:\tilde{X}_{\Delta}\rightarrow X_{\Delta}$ such that the desingularised space $\tilde{X}_{\Delta}$ is smooth enough that any CY variety $\mathcal{M}\subset\tilde{X}_{\Delta}$ can be chosen smooth. 

Consider the particular case where $n=4$ and therefore $\mathcal{M}$ describes a CY threefold.
Since $\mathcal{M}$ has dimension $3$, it can be smoothly deformed around singular loci with codimension $3$ (terminal singularities).
Therefore, we only need to consider desingularisations that resolve everything up to terminal singularities.
If $\tilde{X}_{\Delta}$ contains terminal singularities (\textit{i.e.}, $\tilde{X}_{\Delta}$ is quasi-smooth) we say that $\tilde{X}_{\Delta}$ is $\mathbb{Q}$-factorial. 
Moreover, to ensure the desingularised space $\tilde{X}_{\Delta}$ remains Gorenstein Fano, and therefore projective, the desingularisation must be crepant. 

We define a maximal projective crepant partial (MPCP) desingularisation $\pi:\tilde{X}_{\Delta}\rightarrow X_{\Delta}$ to be one such that the pullback $\pi^{*}$ is crepant, and $\tilde{X}_{\Delta}$ is $\mathbb{Q}$-factorial.
From~\cite{batyrev1993dual,batyrev1994calabiyau}, we have the following theorem:
\begin{theorem}
    Let $X_{\Delta}$ be the toric Gorenstein Fano variety built from a reflexive polytope $\Delta$ with normal fan $\Sigma$.
    Then $\tilde{X}_{\Delta}$ admits at least one MPCP desingularisation
    \begin{equation}
        \pi: \tilde{X}_{\Delta} \rightarrow X_{\Delta} \,.
    \end{equation}
\end{theorem}

Resolution of non-terminal singularities can be seen as refining the open cover $\mathcal{U}(\tilde{X}_{\Delta})$ on $\tilde{X}_{\Delta}$.
Since each cone $\sigma_{\theta}\subset N$ in the normal fan $\Sigma_{\Delta}$ corresponds to a coordinate patch $U\in\mathcal{U}(\tilde{X}_{\Delta})$, refining $\mathcal{U}(\tilde{X}_{\Delta})$ amounts to subdividing the cones $\Sigma_{\Delta}$.
This subdivision corresponds to a triangulation of the dual polytope $\Delta^{*}\subset N$.

To be precise, a \textit{triangulation} of an $n$-dimensional reflexive polytope $\Delta^*$ consists of simplices of dimension $n$ such that the intersection of any two simplices is a face of each and the union of all simplices recovers $\Delta^*$.

A triangulation is said to be:
\begin{itemize}
\item \textit{star} if the origin is a vertex of every full-dimensional simplex.
We require the triangulations of $\Delta^{*}$ to be star so that each subdivision of $\Sigma_{\Delta}$ is still a convex rational polyhedral cone with a vertex at the origin.
\item \textit{fine} if it uses all points in the point configuration.
We require the triangulations of $\Delta^{*}$ to be fine so that all non-terminal singularities are resolved and the resulting CY $\mathcal{M}$ is smooth.
\item \textit{regular} if it can be obtained from the following construction.
Let $\textbf{A}=(\textbf{a}_{1},...,\textbf{a}_{m})$ be a point configuration in $\mathbb{R}^{n}$ and $\Delta$ the convex hull of $\textbf{A}$.
Define a height vector $h=(h_{1},...,h_{m})\in\mathbb{R}^{m}$ and consider the lifted point configuration in $\mathbb{R}^{n+1}$
\begin{equation}
    \textbf{A}^{h}:=
    \begin{pmatrix}
        \textbf{a}_{1} & \cdots & \textbf{a}_{m} \\
        h_{1} & \cdots & h_{m}
    \end{pmatrix} \,.
\end{equation}
Compute the convex hull of $\textbf{A}^{h}$ to obtain the polytope $\Delta^{h}$ and compute the ``lower faces'' of $\Delta^{h}$, where the lower faces are those that have a non-vertical supporting hyperplane $H$ and with $\Delta^{h}$ above $H$.
Projecting down the lower faces of $\Delta^{h}$ to $\mathbb{R}^{n}$ produces a triangulation of $\Delta$.
We require the triangulations of $\Delta^{*}$ to be regular so that $\tilde{X}_{\Delta}$ is projective, and hence Kähler; the CY hypersurfaces inherit this Kähler structure. 
\end{itemize}

The heights generating some regular triangulation are not unique, and in fact, many heights can lead to the same triangulation. Let $h=(h_{1},...,h_{m})$ be the set of heights defining the regular triangulation $\mathcal{T}$ of a polytope $\Delta$.
The set of all height vectors that generate the same triangulation form a polyhedral cone $C$ called the secondary cone.
A simple argument for this is that if $h$ is a set of heights that generates $\mathcal{T}$, then $ch$ also generates $\mathcal{T}$ for any $c>0$ and if $g$ is also a set of heights that generates $\mathcal{T}$ then $h+g$ generates $\mathcal{T}$.

\subsection{Fibration structures}\label{appendix:fibrations}

Given a CY hypersurface in a toric variety $X_{\Sigma}$ associated to a polytope $\Delta \subset N$ with a fan $\Sigma$, the existence of a fibration of form:
\begin{gather}
	X_{\Sigma_{\mathrm{fib}}}\hookrightarrow X_{\Sigma}\stackrel{\pi}{\longrightarrow} B
\end{gather}
over a toric base $B$ with a fan $\Sigma_b$ and $\Sigma_{\mathrm{fib}} \subset \Sigma$ is obstructed by the set of compatibility conditions on the subpolytope $\Delta_{\mathrm{fib}}\subset \Delta$ and the fans $\Sigma, \Sigma_b$ and $\Sigma_\mathrm{fib}$~\cite{Rohsiepe:2005qg}.

In the case when searching for $\mathrm{codim}(\Delta_{\mathrm{fib}} \subset \Delta) = 1$ fibration structures, it is sufficient to show that such $\Delta_{\mathrm{fib}}$ is indeed a reflexive subpolytope~\cite{Rohsiepe:2005qg}. Similarly, in the case when searching for elliptically fibered toric CY threefolds, it has been shown that the existence of a two-dimensional reflexive subpolytope implies existence of compatible fans~\cite{Huang:2019pne}. However, in general, one must ensure that there indeed exists a set of compatible fans $\Sigma$, $\Sigma_b$, and $\Sigma_{\mathrm{fib}}$.

We mainly follow the construction outlined in~\cite{Rohsiepe:2005qg}. In particular, we have:
\begin{definition} Let $\Delta \subset N\simeq \mathbb{Z}^n$ be a reflexive polytope with a fan $\Sigma$. A \textbf{virtual toric $d$-fibration} of $X_\Sigma$ is an embedding $\Delta'\hookrightarrow \Delta$ of a $d$-dimensional reflexive subpolytope $\Delta'$ into $\Delta$.
\end{definition}
Construction of a virtual toric $d$-fibration for a given polytope $\Delta$, assuming one exists, can be formulated as a pure optimization problem by noting that the subpolytope $\Delta'$ can be defined using a set of $d$ vertices of $\Delta$. We shall first briefly describe the construction of $\Delta'$. Let $\{v_0,\dots, v_{k-1}\}$ be the vertices of $\Delta$. Define a state $s$ of the RL environment of virtual $d$-fibrations as a tuple:
\begin{gather}
	s = (n_0, \dots, n_{d-1}) \,,
\end{gather}
where $n_j\in \mathbb{Z}_k^d$ such that $n_j\neq n_k$ for $j\neq k$. Therefore, the state space $\mathcal{S}$ can be identified as a subset of $\mathbb{Z}_k^d$. A subpolytope $\Delta'$ associated to a state $s$ is an intersection of the plane:
\begin{gather}
	H(s) = \bigoplus_{i=0}^{d-1}\mathbb{R}v_{n_i} \,,
\end{gather}
with the polytope $\Delta$, that is: $\Delta' = \Delta \cap H(s)$.

Naturally, we may define an action on the state space $\mathcal{S}$ by the natural group action:
\begin{gather}
	\mathbb{Z}_k^d \times \mathcal{S} \longrightarrow \mathcal{S}
\end{gather}
given by the group product:
\begin{gather}
	(m_0, \dots, m_{d-1})\cdot (n_0,\dots, n_{d-1})	= (m_0 + n_0, \dots, m_{d-1} + n_{d-1}) \,.
\end{gather}
Noting that $\mathbb{Z}_k^d$ is cyclic, we may restrict the set of actions on a given state $s$ to the generators:
\begin{equation}
	a_0 := (1,~0,~\dots,~0)	\,, \quad
	a_1 := (0,~1,~\dots,~0)	\,, \quad 
	\cdots \,, \quad
	a_{d-1} := (0,~0,~\dots,~1) \,,
\end{equation}
which yields the set of actions $\mathcal{A} := \{a_0,\dots,a_{d-1}\}$.

In order to construct the fitness function $f\colon \mathcal{S}\rightarrow \mathbb{R}$, note that there are two main conditions that need to be imposed on the subpolytope $\Delta' = \Delta\cap H(s)$:
\begin{enumerate}
	\item $\Delta'$ must be a reflexive polytope;
	\item $\dim \Delta' = d$.
\end{enumerate}
Although the number of the vertices defining $H(s)$, and hence $\Delta'$, is $d$, it is not necessarily guaranteed that the rank of $\Delta'$ is equal to $d$. The fitness function defined in \eqref{eq:rewardVirtFibrations} penalizes the action on a given state if the resulting state fails to meet any of the conditions above.

\section{Reinforcement learning}\label{sec:RL}

Reinforcement learning is an area of machine learning which works on the principal of trial and error.
An agent performs actions on states in an environment with the goal of maximizing the cumulative reward.
The action changes the state in some way and the agent receives a reward or penalty based on whether the new state is better or worse than the previous one. 
Over time the agent learns which actions to take given the current state in order to maximize the reward. 

The main components of reinforcement learning are:
\begin{itemize}
    \item a set of states $\mathcal{S}$;
    \item a set of actions $\mathcal{A}$; and
    \item a reward function $R(s,a)$ that gives a reward associated to the action $a$ taken on the state $s$. 
\end{itemize}
An agent interacts with the environment in discrete time steps; at each time step $t$ the agent receives the current state $s_{t}$ and a reward $r_{t}$.
The agent then chooses an action from $\mathcal{A}$ and the environment moves to the next state $s_{t+1}$ with reward $r_{t+1}$.
The end goal of the agent is to learn a policy function $\pi:\mathcal{S}\times\mathcal{A}\rightarrow [0,1]$, $\pi(s,a)=\text{Pr}(a_{t}=a|s_{t}=s)$, which maximises the expected cumulative reward.

The state-value function $V_{\pi}(s)$ is defined as the expected discounted return starting at state $s$ and successively following policy $\pi$:
\begin{equation}
    V_{\pi}(s) = \mathbb{E}[G|S_{0}=s] = \mathbb{E} \left[ \sum_{t=0}^{\infty} \gamma^{t}R_{t+1} | S_{0}=s \right] \,,
\end{equation}
where $R_{t+1}$ denoted the reward of moving from state $S_{t}$ to $S_{t+1}$ and $0\leq\gamma<1$ is the discount rate. Since $\gamma<1$ it has the effect of valuing rewards received earlier higher than those received later. $G$ denotes the discounted return, and is defined as the sum of future discounted rewards. 
The agent aims to learn a policy that maximises the expected discounted return. Such a policy is called the optimal policy. 

The action-value of a state-action pair $(s,a)$ under a policy $\pi$ is defined as
\begin{equation}
    Q^{\pi}(s,a) = \mathbb{E}[G|s,a,\pi] \,,
\end{equation}
where $G$ now stands for the random discounted return associated with first taking action $a$ in state $s$ and then following $\pi$ thereafter. 
Then if $\pi^{*}$ is an optimal policy, the optimal action to take from a state $s$ is the one with the highest action-value $Q^{\pi^{*}}(s,\cdot)$. 
The action-value function of such an optimal policy $Q^{\pi^{*}}$ is called the optimal action-value function and is commonly denoted by $Q^{*}$.
Knowledge of the optimal action-value function alone suffices to know how to act optimally. 

\subsection{Q-learning} 

One of the most common reinforcement learning algorithms is Q-learning. Q-learning is based on a $Q$-function that assigns quality for a given action when the environment is in a given state. This value function $Q$ is then iteratively refined. 
C. Watkins first introduced the foundations of Q-learning in his thesis in 1989~\cite{watkins} and gave further details in a 1992 publication titled Q-learning~\cite{Qlearning}.

In Q-learning, the $Q$-function calculates the quality of a state-action combination:
\begin{equation}
    Q: \mathcal{Q}\times\mathcal{A}\rightarrow \mathbb{R} \,.
\end{equation}
Firstly $Q$ is initialised to some constant function. Then at each time step $t$ the agent selects an action $a_{t}$, observes a reward $r_{t}$ and moves to a new state $s_{t+1}$ and $Q$ is updated. 

$Q$ is updated based on the Bellman equation:
\begin{equation}
\label{eq:bellman}
    Q^{new}(s_{t},a_{t}) \leftarrow (1-\alpha)Q(s_{t},a_{t}) + \alpha\left(R_{t+1} + \gamma\max_{a}Q(s_{t+1},a)\right) \,,
\end{equation}
where $R_{t+1}$ is the reward received when moving from state $s_{t}$ to $s_{t+1}$, $0<\alpha<1$ is the learning rate and $0\leq\gamma\leq1$ is the discount factor. 
Therefore, $Q^{new}(s_{t},a_{t})$ is the sum of three factors:
\begin{itemize}
    \item $(1-\alpha)Q(s_{t},a_{t})$: the current value (weighted by one minus the learning rate);
    \item $\alpha R_{t+1}$: the reward if action $a_{t}$ is taken on state $s_{t}$ (weighted by the learning rate); and 
    \item $\alpha\gamma\max_{a}Q(s_{t+1},a))$: the maximum reward that can be obtained from state $s_{t+1}$ (weighted by the learning rate and discount factor).
\end{itemize}

The learning rate $\alpha$ determines to what extent newly acquired information overrides old information.
A factor of $0$ makes the agent learn nothing (exclusively exploiting prior knowledge), while a factor of $1$ makes the agent consider only the most recent information (ignoring prior knowledge to explore possibilities).
In fully deterministic environments, a learning rate of $\alpha_{t}=1$ is optimal.
When the problem is stochastic, the algorithm converges under some technical conditions on the learning rate that require it to decrease to zero.
In practice, often a constant learning rate is used, such as $\alpha_{t}=0.1$ for all $t$. 

In its simplest form the $Q$ function data is stored in a table.
The rows and columns of the table correspond to different states and actions respectively.
The first step is to initialise the Q-table and then as the agent interacts with the environment and receives rewards, the values in the Q-table are updated. 
In cases where the set of states and action is very large the likelihood of the agent visiting a particular state and performing a particular action is small in which case Q-learning can be combined with function approximation.
One solution is to use a deep feed-forward neural network to represent the function.
This gives rise to deep Q-learning.

\bibliographystyle{elsarticle-num} 
\bibliography{references}

\end{document}